\documentclass[preprint,12pt,numbers,sort&compress]{elsarticle}

\usepackage{amsmath,mathtools,amsfonts,stmaryrd,amssymb,dsfont} 
\usepackage{multirow}
\usepackage[hidelinks]{hyperref}

\usepackage[T1]{fontenc} 
\usepackage{textgreek}

\usepackage{hyperref}

\usepackage{xcolor} 
\definecolor{revblue}{RGB}{0,0,150}
\newcommand{\rev}[1]{\textcolor{revblue}{#1}}
\renewcommand{\rev}[1]{#1}

\DeclareSymbolFont{matha}{OML}{txmi}{m}{it}
\DeclareMathSymbol{\varv}{\mathord}{matha}{118}

\journal{Chemical Physics}

\allowdisplaybreaks
\begin{document}

\begin{frontmatter}



\title{Production of Spin-Polarized Molecular Beams via Microwave or Infrared Rotational Excitation}

 \author[kannis]{Chrysovalantis~S.~Kannis}
 \author[rakitzis1,rakitzis2]{T.~Peter~Rakitzis\corref{cor1}}
  
\affiliation[kannis]{organization={Institut f{\"u}r Laser- und Plasmaphysik, Heinrich-Heine-Universit{\"a}t D{\"u}sseldorf},
	addressline={Universit{\"a}tsstra{\ss}e 1},
	city={D{\"u}sseldorf},
	postcode={40225},
	state={NRW},
	country={Germany}}
	
 \affiliation[rakitzis1]{organization={Department of Physics, University of Crete},
             addressline={Voutes},
             city={Heraklion-Crete},
             postcode={70013},
             country={Greece}}

 \affiliation[rakitzis2]{organization={Institute of Electronic Structure and Lasers, Foundation for Research and Technology-Hellas},
             addressline={N.~Plastira 100},
             city={Heraklion-Crete},
             postcode={71110},
             country={Greece}}
\cortext[cor1]{Corresponding author}
\ead{ptr@iesl.forth.gr}


\begin{abstract}
We \rev{propose schemes to produce highly nuclear-spin polarized} small molecules in an intense and cold molecular beam via microwave or infrared rotational excitation, followed by hyperfine-induced quantum beats. Repumping schemes can be used to achieve polarization above $90\%$ in cases where single-pumping schemes are insufficient. We discuss the possibility of high production rates which allow applications including nuclear-magnetic-resonance signal enhancement, and spin-polarized nuclear fusion, where polarized nuclei are known to enhance D-T and D-\textsuperscript{3}He fusion cross sections by $50\%$. 
\end{abstract}

\begin{graphicalabstract}
\includegraphics[width=1.0\columnwidth]{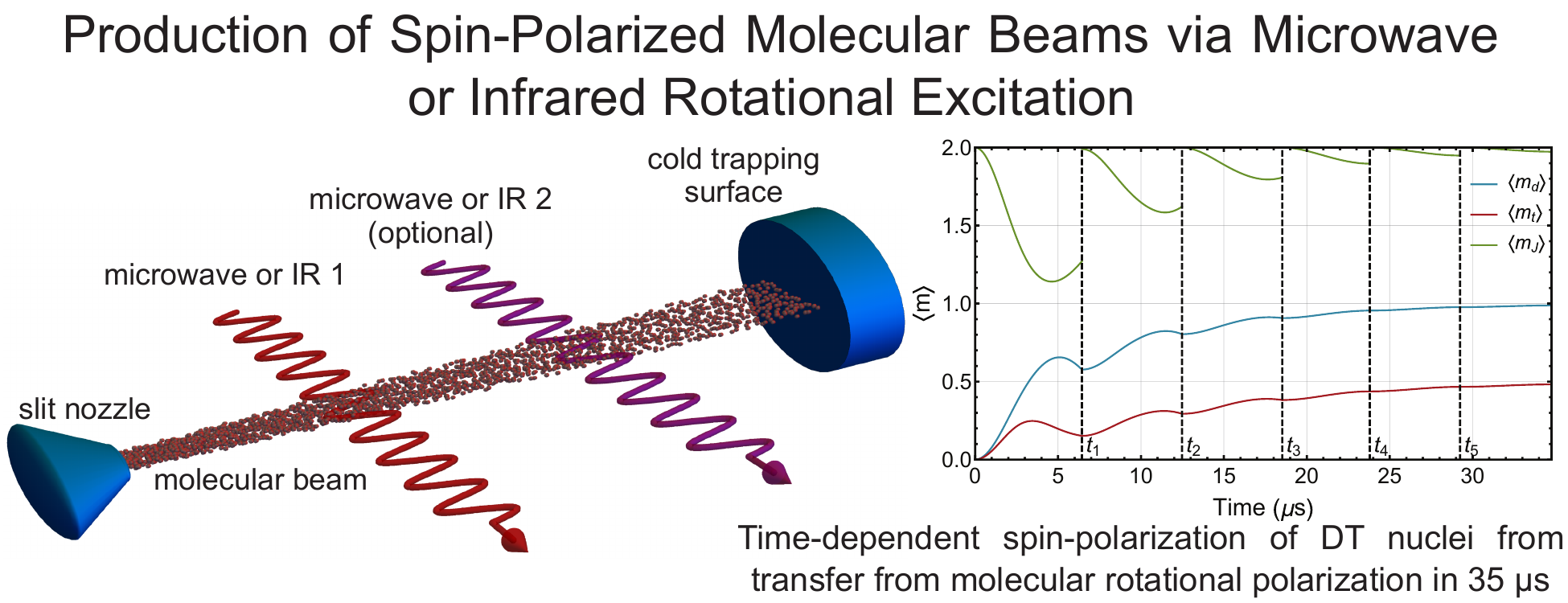}
\end{graphicalabstract}

\begin{highlights}
\item \rev{Schemes proposed to produce spin-polarized molecular beams} with IR\rev{/MW} excitation.
\item \rev{Estimated} production \rev{rates} \rev{may} exceed $10^{21}$~${\text{s}^{-1}}$ from intense IR\rev{/MW} sources.
\item Spin-polarized molecules can enhance NMR/ESR signals or nuclear fusion cross sections.
\end{highlights}


\end{frontmatter}



\section{Introduction}
\label{intro}

The production of spin-polarized molecules is important in several fields, including the study of spin-dependent effects in particle and nuclear physics~\cite{steffens2003}, and in applications of nuclear magnetic resonance  in chemistry and biochemistry~\cite{green2012, keshari2014, levitt2012} and medical resonance imaging (MRI)~\cite{mcrobbie2008}. \rev{Gas phase NMR spectroscopy~\cite{jackowski2016,jackowski2025} provides accurate molecular spectral parameters free from intermolecular interaction effects and therefore serves as a direct benchmark for quantum-chemical calculations of isolated molecules. To the best of our knowledge, direct detection of NMR signals from molecules whose nuclear spin polarization is created via rovibrational excitation in the gas phase has not yet been reported.} In all cases, the higher the spin polarization, the higher the desired signal. The spin-polarization $P$ induced by a magnetic field at room temperature (for example, in medical imaging) is usually limited to no more than $P{\sim}10^{-5}$, therefore, spin-polarized molecules with polarizations near unity can enhance MRI signals by several orders of magnitude. Techniques that significantly extend the hyperpolarization lifetime~\cite{burueva2017} enable further applications in MRI studies. Furthermore, it is known that complete nuclear polarization increases the fusion cross section for the D-T and D-\textsuperscript{3}He reactions by ${\sim} 50\%$~\cite{hupin2019}, and may increase the reactor efficiency by ${\sim} 75\%$~\cite{temporal2012}. Additionally, it was predicted recently that spin-polarized D and T nuclei can reduce the minimum startup tritium inventory by more than an order of magnitude~\cite{parisi2024}. Finally, a first demonstration of polarization persistence in a laser-induced plasma was performed very recently~\cite{zheng2024}, which is an important step for showing that spin polarization can survive in the plasma on timescales sufficient to benefit laser-induced fusion reactions. Multiple studies~\cite{bae2025, garcia2025, bakri2025, reichwein2024, heidbrink2024, hu2023,baylor2023,cook2025} further highlight the potential advantages of spin-polarized fuel in both magnetic and inertial confinement fusion, underscoring the growing interest in this direction.

However, methods for producing highly spin-polarized molecules are limited both by the quantity of molecules produced and by their purity, needed for magnetic resonance enhancement or polarized fusion applications. Dynamic nuclear polarization (DNP) and cryogenic cooling are the only methods that are routinely used to produce spin-polarized molecules in macroscopic quantities. DNP uses highly-polarized unpaired electrons in free radicals at low temperature in a magnetic field, to transfer this polarization to the nuclear spins of the target sample~\cite{ardenkjaerlarsen2003}. However, the free radicals need to be removed for medical applications, which is a difficult step that limits the applicability of the technique. Similarly, cryogenic cooling has been used to produce spin-polarized D\textsubscript{2} at a polarization of about $20\%$. The total production rates and polarizations are sufficient for proof-of-principle experiments, but otherwise are too small for the requirement of a nuclear fusion reactor: a ${\sim}1$ GW nuclear reactor would require production rates of at least ${\sim}10^{21}$~${\text{s}^{-1}}$~\cite{grigoryev2011, kulsrud1987, moir1992}.

The Stern-Gerlach spin-separation technique, and spin-exchange optical pumping, are limited to microscopic production rates of ${\sim}10^{17}$~${\text{s}^{-1}}$ (${\sim}1$~$\text{\textmu mol}\,{\text{s}^{-1}}$) for atoms~\cite{nass2003, clasie2006}, and many orders of magnitude lower for molecules~\cite{kravchuk2011, horke2014}. These microscopic production rates are small and expensive for medical applications, and insufficient for nuclear fusion applications. 

An alternative method for the production of spin-polarized molecules from the infrared (IR) excitation of molecular beams~\cite{rakitzis2005, rubiolago2006} has been demonstrated~\cite{sofikitis2007, bartlett2008, bartlett2009, bartlett2010}. In this method, the rotational angular momentum of the molecules is polarized in a rovibrational transition, and this polarization is transferred to the nuclear spin via the hyperfine interaction, on the timescale of about $10$--$100$~$\text{\textmu s}$ (for spin-$1/2$ nuclei; higher spins can be polarized more quickly). The polarization can then be frozen in the molecular nuclei by stopping the spin-rotation interaction, either by photodissociating the molecule, turning on a magnetic field, or freezing the molecule at a surface (and stopping the molecular rotation).

This IR-excitation method can then produce orders-of-magnitude more spin-polarized molecules than the Stern-Gerlach spin-separation beam technique, for the following reasons. Both these beam techniques have production rates given by the product of the beam parameters $A \rho v$, where $A$ is the cross-sectional area, $\rho$ is the molecule number density, and $v$ is the beam velocity. For the Stern-Gerlach method, the largest spin-polarized H-atom production rate of $A \rho v \sim 5 \times 10^{16}$~$\text{H}/\text{s}$ was achieved with $A\sim 1$~$\text{cm}^2$, $\rho\sim 2\times 10^{11}$~$\text{H}/\text{cm}^3$, and $v\sim 2.5\times 10^{5}$~$\text{cm}/\text{s}$~\cite{hupin2019}. The beam density $\rho$ or area $A$ cannot be increased, without degrading the polarization from diminished spin separation, and $v$ is determined by the supersonic beam expansion. For spin separation of molecules, the densities are about $3$ orders of magnitude lower, as the nuclear magnetic moments are about $3$ orders smaller than those of electrons. In contrast, the IR excitation method does not require spin separation, as polarization transfer through hyperfine beating occurs within each molecule, on a timescale $3$--$4$ orders of magnitude faster than spin separation. Therefore, the beam densities $\rho$ can be at least $3$--$4$ orders higher, and the beam area $A$ can be $1$--$2$ orders larger, allowing beam production rates of ${\sim} 10^{22}$~$\text{s}^{-1}$. Recently, the authors have proposed the IR excitation of such formaldehyde (CH\textsubscript{2}O) and formic acid (CH\textsubscript{2}O\textsubscript{2}) beams for the production of spin-polarized H\textsubscript{2} isotopes~\cite{kannis2021}, at production rates of ${\sim} 10^{20}$~$\text{s}^{-1}$, limited by commercially available tabletop IR lasers with photon production rates of ${\sim} 10^{21}$~$\text{s}^{-1}$.

\rev{We present a theoretical analysis of IR or microwave excitation schemes which, according to our numerical simulations, can lead to high nuclear spin polarization of a molecular beam and efficient collection of the polarized particles under realistic conditions. The proposed schemes are designed to polarize essentially the full population of the beam and hence enable collection and storage without additional state-selection steps. In contrast to earlier approaches~\cite{kannis2021, sofikitis2015}, which relied on photodissociation either for the production and collection of polarized photofragments~\cite{kannis2021} or for the removal of undesired spin states~\cite{sofikitis2015}, the present schemes avoid such secondary processes. This reduction in experimental steps may simplify implementation and improve overall production efficiency.}


The paper is organized as follows. In Section~\ref{method}, we describe the molecular-beam excitation method in detail. In Section~\ref{sec3}, we show the numerical simulations and analyze the resulting polarization dynamics for small molecules, including HD, DT, O\textsubscript{2}, NO, N\textsubscript{2}O, CO, and H\textsubscript{2}S\textsubscript{2}. Finally, Section~\ref{conclusion} summarizes the main conclusions.

\section{Description of the method}
\label{method}

Here, we present a proposal for a general method for the production of spin-polarized small molecules, through the rotational excitation of molecular beams with microwaves. This method promises significantly higher production rates of spin-polarized molecules, as $10$~W of microwaves in the $15$ to $150$~GHz range corresponds to polarized photon production rates of $10^{23}$--$10^{24}$~$\text{s}^{-1}$, which is the limiting rate for spin-polarized molecule production. These rates are sufficient for medical-imaging and polarized-fusion applications. We describe methods for the polarization of isotopes of several small molecules: HD, DT, O\textsubscript{2}, NO, N\textsubscript{2}O, CO, and H\textsubscript{2}S\textsubscript{2}. The first six will be trapped directly after polarization onto a cold surface, whereas H\textsubscript{2}S\textsubscript{2} is unstable and will be dissociated to yield and trap spin-polarized H\textsubscript{2} isotopes.

\begin{figure}
	\centering
	\includegraphics[width=1.0\columnwidth]{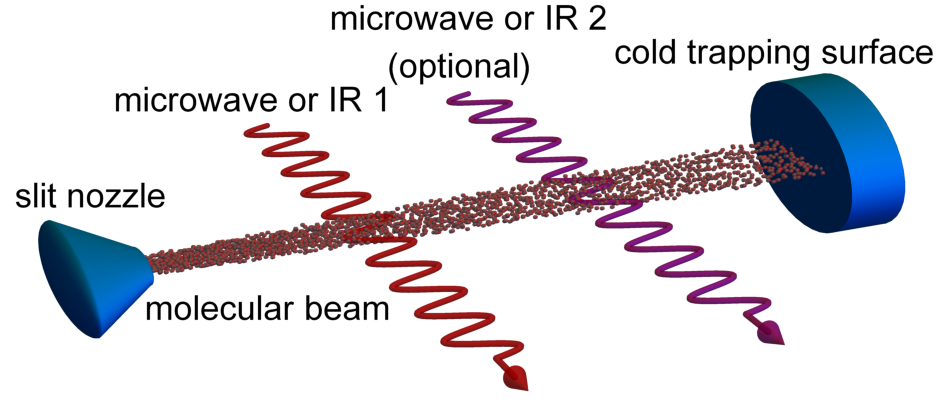}
	\caption{Sketch of the experimental setup. A cool molecular beam is excited to the state $\left\vert \varv \, J \, m_{J} \right\rangle$---i.e., rotationally polarized---via microwave or IR radiation. This polarization is subsequently transferred to the nuclear spin. Additional excitation and polarization-transfer steps may be required. Finally, the polarized beam is collected on a cold surface.}\label{fig1}
\end{figure}

We note that the ability to rotationally excite some of these molecules near the $150$~GHz range, or the rovibrational excitation of HD and DT with tunable narrowband infrared light, in macroscopic quantities, has been improved by recent improvements in microwave amplifies and IR laser sources. Furthermore, we employ Stimulated Raman adiabatic passage (STIRAP~\cite{bergmann2019}) for IR excitation schemes, or \textpi-pulses for microwave excitation schemes, that allow $100\%$ population transfer from the ground state to the target excited state. The particular schemes utilized depend on the nuclear spins and rotational constants of the molecule\rev{~\cite{brown2003,sukhenko1985}}. We first give a general description of the excitation schemes of molecular beams (as shown in Fig.~\ref{fig1}), before giving specific details for the molecules HD, DT, O\textsubscript{2}, NO, N\textsubscript{2}O, CO, and H\textsubscript{2}S\textsubscript{2}: 
\begin{enumerate}
	\item Production of a molecular beam expansion, with a production rate of up to ${\sim}10^{23}$~$\text{s}^{-1}$, which cools the rotational population to predominantly the lowest molecular rotational state ($J=0$)~\cite{grieser2018,schlimme2021}.
	\item Circularly polarized microwave or IR beams, transfer ideally $100\%$ of the molecules from the ground state to specific rovibrational states $\left\vert \varv \, J \, m_{J} \right\rangle$, from which subsequently the rotational polarization is transferred to the nuclear spin. In some cases more than one excitation and polarization-transfer step is needed.
	\item After the nuclear spins are polarized, the hyperfine beating can be stopped, when the molecular beam enters a region with a sufficiently large magnetic field~\cite{engels2018,kannis2018}.
	\item The polarized molecular beam is finally collected and stored at a cold surface~\cite{govers1980, engels2015}.
\end{enumerate}
In this way, macroscopic production rates of spin-polarized HD, DT, O\textsubscript{2}, NO, N\textsubscript{2}O, and CO molecules can be achieved in excess of ${\sim} 10^{21}$~$\text{s}^{-1}$.

All these steps have been demonstrated individually for different molecules, as supported by the references provided. Some typical examples include: (1) The nozzle for producing a cooled supersonic beam expansion has been operated with H\textsubscript{2}, Ar, O\textsubscript{2}, N\textsubscript{2}, etc.~\cite{grieser2018}. (2) STIRAP has been used to prepare oriented and aligned samples of H\textsubscript{2} ($\varv=1, \, J=1, 2, 3$) and HD ($\varv=1, \, J=2$)~\cite{bartlett2008}. Under optimized conditions, transfer efficiencies of $\sim 100\%$ are achievable~\cite{vitanov2017}. The resulting rotational depolarization, or equivalently the induced nuclear spin polarization via hyperfine interactions, has been demonstrated for HD and D\textsubscript{2} in~\cite{bartlett2009} and~\cite{bartlett2010}, respectively. (3) The preservation of polarization in H\textsubscript{2}, D\textsubscript{2}, and HD has been demonstrated by applying an external magnetic field stronger than a critical threshold~\cite{engels2018}. (4) Sticking of unpolarized H\textsubscript{2} and D\textsubscript{2} beams to a low-temperature substrate has been achieved experimentally~\cite{govers1980}. Rotational and vibrational relaxation are expected for molecules that are initially in excited states, as they release energy and settle into lower-energy configurations upon sticking to the cold surface. Molecules already in the rovibrational ground state, such as HD (or DT), are unaffected by these processes.

\rev{For realistic cold beam jet parameters~\cite{vestrick2024}, longitudinal velocity dispersion ($\delta v /v = 1\%$) introduces only minor corrections to the pulse area and detuning and therefore does not impose fundamental restrictions on the proposed coherent excitation schemes. Specifically, for $\pi$-pulses (for example, at a typical frequency of $100$~GHz) with a duration of $\tau_{p} = 10~\text{\textmu{s}}$, the Doppler shift is more than an order of magnitude smaller than the Rabi frequency, while the spectral width of such a pulse ($1/\tau_{p}$), also exceeds the Doppler broadening by more than an order of magnitude. Similarly, for the STIRAP scheme, the influence of velocity spread is discussed at the end of Sec.~\ref{sechd}.}

Spin relaxation may occur due to local magnetic field fluctuations, magnetic dipole–dipole interactions between neighboring molecules, or quadrupole interactions if the nuclei involved have spin greater than $1/2$. However, applying a sufficiently strong external magnetic field can help suppress these spin relaxation mechanisms and preserve nuclear polarization. Spin-polarized HD molecules have been trapped for months, without significant degradation of the polarization~\cite{bass2014,lowry2016}, showing that spin-polarized HD can be successfully stored for long periods of time.

The main issue at the surface is removing the heating from the kinetic energy of the molecular beam trapped at the surface. For $10^{21}$~$\text{s}^{-1}$, and a beam speed of about $1000$~$\text{m}\,\text{s}^{-1}$, this comes out to $2.5$~W, which is close to the cooling limit of commercial Pulse-Tube Cryocoolers. For $10^{22}$~$\text{s}^{-1}$ production rates (needed for a nuclear reactor), a custom cooling design will likely be necessary. However, commercial Cryocoolers will be sufficient for proof-of-principle setups. Further mitigation of this problem may be to have the HD beam speed as low as $300$~$\text{m}\,\text{s}^{-1}$, by expanding with Ar carrier gas, which will reduce the kinetic energy of the HD beam by a factor of $10$, and reduce the need for surface cooling by the same factor.

\rev{The experimental realization of the polarization scheme shown in Fig.~\ref{fig1} would ideally employ a weak and highly homogeneous magnetic field generated by electromagnets inside \textmu-metal shielding. Such shielding suppresses external stray fields, including the Earth's field and those produced by nearby equipment. Under these conditions a well-defined quantization axis can be established and controlled, which is advantageous both for preparing the polarization and preserving it until detection. The quantization axis is assumed to be parallel to the molecular beam direction and defines the axis with respect to which the spin and angular momentum projections are specified. The electric field of circularly polarized laser beams is perpendicular to both the molecular beam and the laser propagation direction, whereas that of a linearly polarized beam is aligned with the magnetic-field axis. In the following, however, the spin dynamics are evaluated in the limit of vanishing magnetic field, since the field required for defining the quantization axis is assumed to be sufficiently weak that it does not significantly influence the internal dynamics.}

\section{Spin-polarization dynamics in HD, DT, O\textsubscript{2}, NO, N\textsubscript{2}O, \textsuperscript{13}CO, and H\textsubscript{2}S\textsubscript{2}}
\label{sec3}

\subsection{HD and DT}
\label{sechd}

\begin{figure*}
	\includegraphics[width=1.0\textwidth]{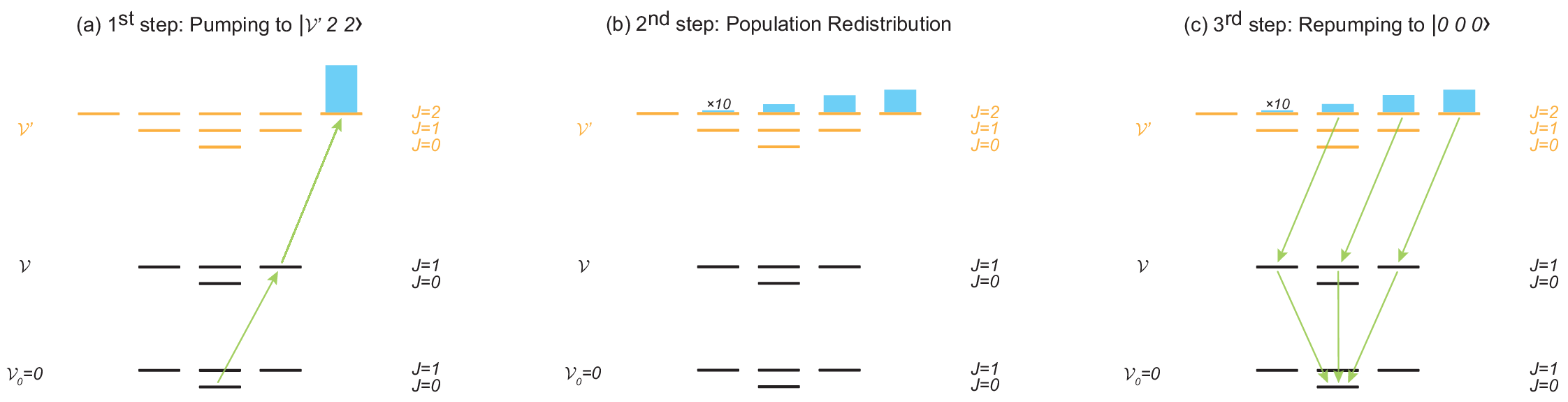}
	\caption{Excitation-polarization scheme for HD or DT via a $J=2$ state. 
		(a) Initial excitation from the lowest rovibrational state to $\left\vert \varv^{\prime}\, 2\, 2 \right\rangle$ at $t=0$.
		(b) Redistribution of population among all $m_{J}$ states until $t=t_{1}$, when the $m_{J}=-1$ population reaches a \rev{local} minimum.
		(c) At $t=t_{1}$, the $m_{J}=0,1,2$ states are selectively repumped to the ground state, preserving the developed nuclear polarization.}\label{figadd1}
\end{figure*}

Figure~\ref{figadd1} shows the excitation scheme adopted for the HD and DT molecules, utilizing an excited rovibrational state with $J=2$. First, the system is pumped from the ground state $\left\vert 0\, 0\, 0\right\rangle$ to the excited rovibrational state $\left\vert \varv^{\prime} \, 2\, 2\right\rangle$ using two right circularly polarized laser beams. Due to hyperfine interactions, the population of this state redistributes among all $m_{J}$ states, leading to the emergence of nuclear polarization. At time $t=t_{1}$, when the population of the $m_{J}=-1$ state reaches a \rev{local} minimum, the $m_{J}=0,1,2$ states are repumped to the ground state, as depicted in Fig.~\ref{figadd1}(c). This ground state retains the induced nuclear polarization, as it does not exhibit hyperfine beating. Finally, the molecules can adhere to the cold surface, having no residual internal energy.

\rev{As a representative case of the polarization scheme, we consider the excited vibrational states $\varv{ = }1$ and $\varv^{\prime} {=} 2$. For the transition $\left\vert \varv_{0} =0 \, J = 0\, m_{J} = 0 \right\rangle \rightarrow \left\vert \varv = 1 \, J = 1\, m_{J} = 1 \right\rangle \rightarrow \left\vert \varv^{\prime} = 2 \, J = 2\, m_{J} = 2 \right\rangle$ shown in Fig.~\ref{figadd1}(a), the radiation fields coupling the first and second states and the second and third states are referred to as pump and Stokes fields, respectively. A counterintuitive pulse sequence, in which the Stokes field is applied before the pump field, can lead to efficient population transfer. The delay between the two fields is realized experimentally as a spatial displacement between the laser beams. This displacement must be optimized depending on the experimental conditions, but the two laser beams should partially overlap. A similar counterintuitive sequence is applied for the deexcitation process shown in Fig.~\ref{figadd1}(c), where the applied fields induce stimulated emission. The corresponding transitions are electric dipole (E1) transitions. For HD, the wavelengths are $2.69~\text{\textmu{m}}$ (experimental)~\cite{fast2020} for $\left\vert \varv_{0} =0 \, J = 0 \right\rangle \leftrightarrow \left\vert \varv = 1 \, J = 1 \right\rangle$ and $2.77~\text{\textmu{m}}$ (theoretical)~\cite{h2spectre, komasa2019} for $\left\vert \varv = 1 \, J = 1 \right\rangle \leftrightarrow \left\vert \varv^{\prime} = 2 \, J = 2 \right\rangle$. For DT, the corresponding wavelengths are $3.58~\text{\textmu{m}}$(theoretical) and $3.66~\text{\textmu{m}}$(theoretical)~\cite{h2spectre, komasa2019}.}

The transfer of rotational angular momentum polarization to the nuclear spins in HD and DT molecules is enabled through hyperfine interactions, which are described by the effective hyperfine Hamiltonian~\cite{ramsey1957,jozwiak2020,jozwiak2020b}:
\begin{equation}\label{eq:hd}
	\begin{split}
		H/h = & -c_{p/t} \mathbf{I_{p/t}} \cdot \mathbf{J} - c_{d} \mathbf{I_{d}} \cdot \mathbf{J}  +\frac{5 d_{1}}{(2 J -1)(2 J + 3)} \\
		& \times \Big[\frac{3}{2} (\mathbf{I_{p/t}} \cdot \mathbf{J})(\mathbf{I_{d}} \cdot \mathbf{J}) +\frac{3}{2} (\mathbf{I_{d}} \cdot \mathbf{J})(\mathbf{I_{p/t}} \cdot \mathbf{J}) \\
		& - \mathbf{I_{p/t}} \cdot \mathbf{I_{d}} \mathbf{J}^2 \Big] +\frac{5 d_{2}}{(2 J -1)(2 J + 3)} \\
		& \times \Big[3 (\mathbf{I_{d}} \cdot \mathbf{J})^2 + \frac{3}{2} (\mathbf{I_{d}} \cdot \mathbf{J}) - \mathbf{I_{d}}^2 \mathbf{J}^2 \Big],
	\end{split}
\end{equation}
where $H/h$ is given in frequency units. The rotational angular momentum operator $\mathbf{J}$ as well as the proton/triton ($\mathbf{I_{p/t}}$) and deuteron ($\mathbf{I_{d}}$) spin operators are considered dimensionless. The first two terms represent the nuclear spin-rotation interactions. The third term corresponds to the direct spin-spin magnetic interaction between the two nuclei, while the fourth term accounts for the interaction of the electric quadrupole moment  of the deuteron.

The theoretical values of hyperfine constants \rev{for the excited vibrational state $\varv^{\prime} = 2$} are~\cite{jozwiak2020,jozwiak2020b}: $c_{p} = 82.32$~kHz, $c_{d} = 12.64$~kHz, $d_{1} = 16.521$~kHz, $d_{2} = -21.793$~kHz for HD, and $c_{t} = 49.734$~kHz, $c_{d} = 7.158$~kHz, $d_{1} = 18.095$~kHz, $d_{2} = -22.135$~kHz for DT\rev{\footnote{\rev{The electron coupled spin-spin interaction term proportional to $\mathbf{I_{d}} \cdot \mathbf{I_{p/t}}$ was neglected in the work providing the effective hyperfine Hamiltonian of Eq.~(\ref{eq:hd}) with the corresponding constants, and is therefore omitted in the present treatment.}}}. The matrix representation of the hyperfine Hamiltonian can be readily obtained in the uncoupled basis $\left\vert J \, m_{J} , I_{d}\, m_{d} , I_{p} \, m_{p} \right\rangle = \left\vert J \, m_{J} \right\rangle \otimes \left\vert I_{d}\, m_{d} \right\rangle \otimes \left\vert  I_{p} \, m_{p} \right\rangle$, thanks to its explicit structure (Eq.~(\ref{eq:hd})). Since in our case $J=2$, $I_{d}=1$, and $I_{p}=1/2$ are fixed, we simplify the notation to $\left\vert m_{J} , m_{d} , m_{p}\right\rangle$. This yields $30$ state vectors which are employed for describing the system's spin state. Furthermore, we assume that during each optical pumping step, the radiation field affects only the rotational angular momentum, leaving the nuclear spin projections unchanged. For instance, when the system is initially pumped into the $\left\vert \varv^{\prime} = 2 \, J=2 \, m_{J} =2 \right\rangle$ state, its initial hyperfine state is taken to be a statistical mixture of the six equally weighted states: $\left\vert m_{J} = +2 , m_{d} =0,\pm 1 , m_{p} = \pm 1/2\right\rangle$. Diagonalizing the Hamiltonian matrix provides the eigenenergies and facilitates the construction of the unitary time-evolution operator, $U=e^{-i H t/\hbar}$, used to calculate the time-evolved state vectors $\left\vert m_{J} , m_{d} , m_{p}\right\rangle_{t}$ at any given time $t$ (for more details, see~\cite{kannis2018}). This approach is valid as long as the Hamiltonian is static. If this is not the case, the formalism described in Ref.~\cite{kannis2025} can be employed.

From these, we can compute transition probabilities between initial and final states and evaluate the expectation values of the $z$-components of the relevant angular momentum operators--$J_{z}$, $I_{d,z}$, and $I_{p,z}$. As these operators are considered dimensionless and satisfy, e.g., $J_{z} \left\vert m_{J} , m_{d} , m_{p}\right\rangle = m_{J} \left\vert m_{J} , m_{d} , m_{p}\right\rangle$ (similarly for $I_{d,z}$ and $I_{p,z}$), we simplify the notation by denoting the corresponding expectation values as $\left\langle m_{J}\right\rangle$, $\left\langle m_{d}\right\rangle$, and $\left\langle m_{p/t}\right\rangle$. Further details and explicit definitions regarding this framework can be found elsewhere~\cite{kannis2018}.  

A single polarization cycle can produce moderate nuclear polarization. However, by repeating this cycle multiple times, it is possible to achieve high nuclear polarization with effectively no population loss. In practice, the same set of beams and polarizations can intersect the molecular beam at successive positions along its propagation axis, corresponding to the time delay between excitation steps. The IR or microwave setup can thus be reused at different locations to implement additional cycles, without requiring a proportional increase in the number of independent laser beams. For a full cycle (pumping up from $\varv=0$, $J=0$ and then eventually back down to $\varv=0$, $J=0$), no photons are used up (ideally); the laser beams merely need to be repolarized. A cycle is considered completed after the selective repumping of the $m_{J}=0,1,2$ state populations to the ground state, i.e., upon completion of the third step in Fig.~\ref{figadd1}. The cycle duration is taken to be equal to the free evolution period during the second step of Fig.~\ref{figadd1}. The next cycle begins immediately after the previous one ends, with no time delay. While a delay could easily be incorporated into the model, it is set to zero in this example. Furthermore, polarization damping between cycles is neglected, and the laser field is assumed to have no direct effect on the nuclear spins.

\begin{table*}
	\centering
	\resizebox{\textwidth}{!}{%
		\begin{tabular}{c c c c c c c c c}
			\hline\hline
			\multirow{2}{*}{\textbf{Cycle}} & \multicolumn{2}{c}{\textbf{Duration}} & \multicolumn{2}{c}{\textbf{Deuteron Polarization}}  & \multicolumn{2}{c}{\textbf{Proton/Triton Polarization}} & \multicolumn{2}{c}{\textbf{Population Loss}} \\
			& HD & DT & HD & DT & HD & DT & HD & DT \\
			\hline
			$1$ & $4.52$~{\textmu s} & $6.44$~{\textmu s} & $59.4\%$ & $57.4\%$ & $20.0\%$ & $30.1\%$ & $0.3\%$ & $0.3\%$ \\
			$2$ & $4.52$~{\textmu s} & $6.04$~{\textmu s} & $77.1\%$ & $80.4\%$ & $46.9\%$ & $58.7\%$ & $0.07\%$ & $0.03\%$ \\
			$3$ & $7.85$~{\textmu s} & $6.04$~{\textmu s} & $86.9\%$ & $90.7\%$ & $72.1\%$ & $76.4\%$ & $0.01\%$ & $0.002\%$ \\
			$4$ & $4.56$~{\textmu s} & $5.28$~{\textmu s} & $92.6\%$ & $95.5\%$ & $80.3\%$ & $87.4\%$ & $0.0007\%$ & $0.0005\%$ \\
			$5$ & $3.00$~{\textmu s} & $5.44$~{\textmu s} & $94.7\%$ & $97.7\%$ & $91.8\%$ & $93.2\%$ & $0.002\%$ & $0.0001\%$ \\
			$6$ & $2.84$~{\textmu s} & $5.52$~{\textmu s} & $96.4\%$ & $98.8\%$ & $96.2\%$ & $96.4\%$ & $0.0008\%$ & $0.00002\%$ \\
			\hline\hline
		\end{tabular}}
	\caption{Parameters of successive polarization cycles in HD and DT.}\label{tabadd1}
\end{table*}

Table~\ref{tabadd1} summarizes the key parameters characterizing each polarization cycle in HD and DT. The duration of each cycle is tailored to its specific objective. In particular, the first four cycles in HD are timed to minimize population loss from the $\left\vert 2 \, 2\, -1\right\rangle$ state, while the final two cycles are optimized to further enhance proton polarization--by which point population losses have already been reduced to a negligible level. In the case of DT, the first three cycles serve to reduce population loss, and the remaining three are dedicated to maximizing triton polarization. After six polarization cycles, the nuclear polarization in both molecules exceeds $96\%$, with minimal population loss. Alternative timing strategies may be employed depending on the desired final polarization.

\begin{figure*}
	\includegraphics[width=1.0\textwidth]{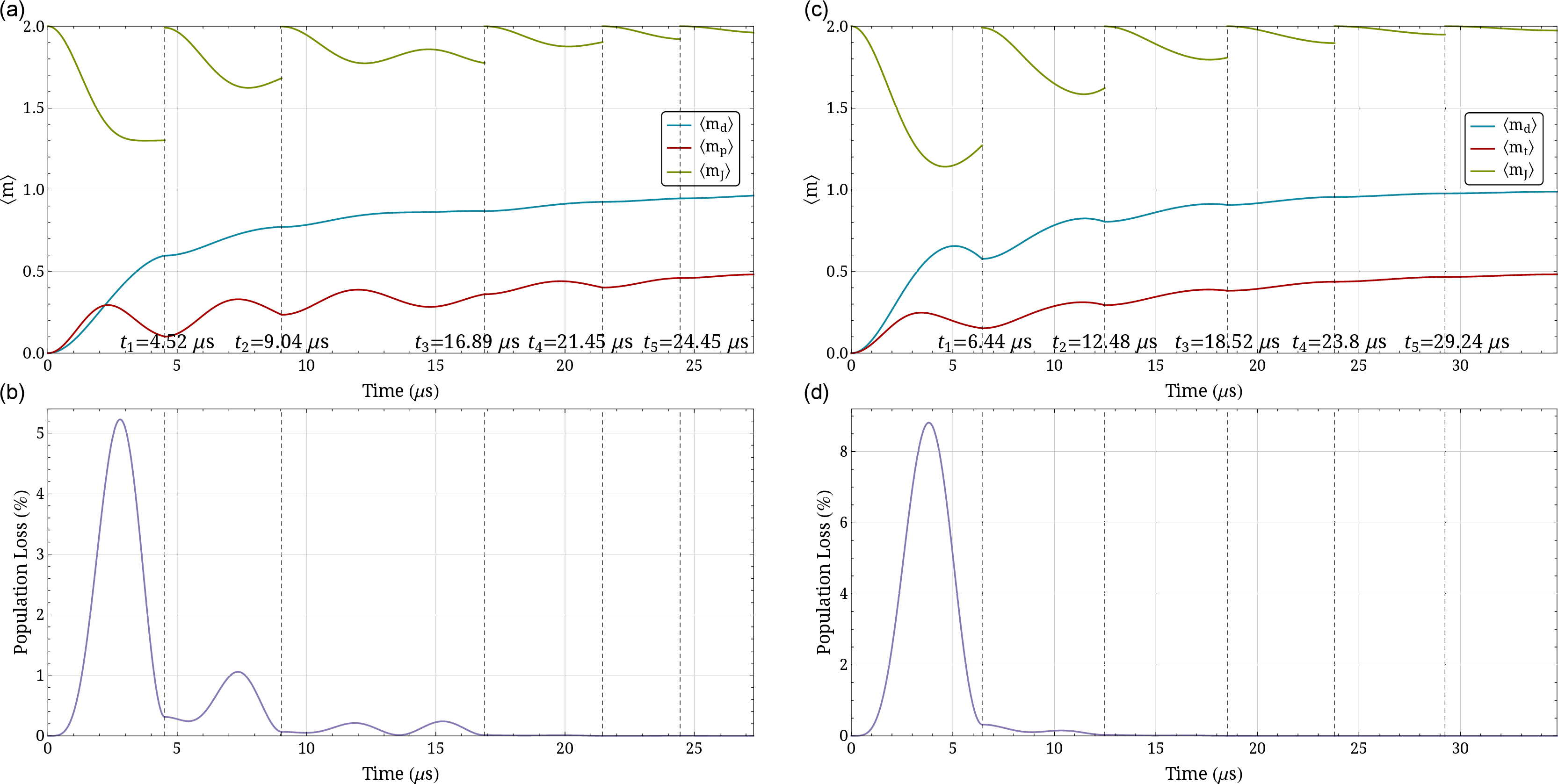}
	\caption{Time evolution of expectation values of the rotational angular momentum projection $\langle m_{J}\rangle$ and nuclear spin projections $\langle m_{d}\rangle$ and $\langle m_{p/t}\rangle$ in panels (a) and (c), and of population loss in panels (b) and (d), over six polarization cycles in HD and DT. The plots correspond to the  $\left\vert \varv^{\prime} =2 \, J=2\right\rangle$ excited state, and therefore the slight reduction in nuclear polarization due to the repumping process at the end of each cycle is not depicted.}\label{figadd2}
\end{figure*}

Figure~\ref{figadd2} illustrates the time evolution of the expectation values of the nuclear spin projections $\langle m_{d}\rangle$ and $\langle m_{p/t}\rangle$ along with the population loss, over six polarization cycles. It is important to note that these values reflect the excited-state properties at the end of the second step in each cycle (see Fig.~\ref{figadd1}(b)), i.e., just before repumping to the ground state. The corresponding nuclear polarization values after repumping--i.e., at the end of each complete cycle--are listed in Tab.~\ref{tabadd1} and are slightly lower than those shown in the figure.

In a realistic experimental implementation of the proposed polarization scheme, the molecular beam exhibits a finite velocity spread. This affects the excitation and de-excitation efficiencies and also leads to smearing of the hyperfine beatings shown in Fig.~\ref{figadd2}. In STIRAP, the relevant effect of the velocity spread arises from the Doppler shift, which contributes to the effective detuning through the velocity component along the laser propagation direction. For counterpropagating pump and Stokes laser beams in a ladder linkage, however, this contribution is strongly reduced: the residual Doppler detuning scales as $\lvert k_{P}-k_{S}\rvert v_{k}$~\cite{vitanov2017}. As a result, when the two photons have similar wavelengths the velocity-induced detuning is negligible, and even for unequal wavelengths the full velocity distribution can be efficiently addressed by operating in the adiabatic regime at sufficiently high laser intensities. The smearing of the hyperfine beatings is likewise negligibly small. For a cool molecular beam with a Gaussian velocity distribution of $1\%$ standard deviation~\cite{vestrick2024}, the expectation values shown in Fig.~\ref{figadd2} vary only at the level of a fraction of a percent. Overall, the velocity distribution of the molecular beam does not significantly degrade the performance of the polarization scheme compared to the idealized case.

\subsection{O\textsubscript{2}}

Next, we consider the case of molecular oxygen O\textsubscript{2}, an open-shell molecule with two unpaired electrons. In particular, its electronic ground state $^{3}\Sigma^{-}_{g}$ has been extensively studied with microwave spectroscopy techniques. This state has a total electron spin of $S=1$ and orbital angular momentum projection of $\Lambda = 0$. The angular momentum quantum number for rotation is denoted by $N$, which together with the electron spin, forms the total angular momentum $J$. Due to symmetry constraints $N$ can only take odd integer values. Within the $N=1$ rotational level \rev{and the ground vibrational state}, we consider the excitation $\left\vert J = 0\, m_{J} = 0 \right\rangle \rightarrow \left\vert J = 1 \, m_{J} = 1 \right\rangle$ \rev{ corresponding to a magnetic dipole (M1) transition at $2.52$~mm~\cite{yu2014}}. This results in a $50\%$ polarization for the total electron spin and $50\%$ rotational polarization, both of which remain time-independent. A subsequent transition, $\left\vert 1 \,  1 \right\rangle \rightarrow \left\vert 2 \,  2 \right\rangle$\rev{, also a magnetic dipole (M1) transition at $5.33$~mm~\cite{yu2014}}, can achieve an electron polarization close to $100\%$. Specifically, the electron polarization oscillates between $89.8\%$ and $100\%$ with a period of $2.1$~ps. The time-averaged electron spin polarization is approximately $95\%$.

The fine-structure effective Hamiltonian matrix elements (in frequency units) are explicitly derived in Ref.~\cite{larsson2019}. Below we summarize them, focusing only on the variation in $N$:
\begin{align}
	\begin{split}
		H^{N{=}J} {=} & \left\langle N{=}J | H | N{=}J \right\rangle {=} B J (J+1) - D J^{2} (J+1)^{2} \\
		& + H J^{3} (J+1)^3 -\gamma- \gamma_{D} J (J+1) \\
		& - \gamma_{H} J^{2} (J+1)^{2} +\frac{2}{3} \big[\lambda + \lambda_{D} J (J+1) \\
		& + \lambda_{H} J^{2} (J+1)^{2}\big]
	\end{split}
	\\
	\begin{split}
		H^{J-1, J-1} {=} & \left\langle N{=}J-1 | H | N{=}J-1 \right\rangle {=} B J (J-1) \\
		& - D J^{2} (J-1)^{2} + H J^{3} (J-1)^3 +\big[\gamma \\
		& +\gamma_{D} J (J-1) + \gamma_{H} J^{2} (J-1)^{2}\big](J-1) \\
		& + \big[\lambda + \lambda_{D} J (J-1) + \lambda_{H} J^{2} (J-1)^{2}\big]\\
		& \times \Bigg(\frac{2}{3}-\frac{2 J}{2 J+1}\Bigg)
	\end{split}
	\\
	\begin{split}
		H^{J+1, J+1} {=} & \left\langle N{=}J+1 | H | N{=}J+1 \right\rangle {=} B (J+2) (J+1) \\
		& - D (J+2)^{2} (J+1)^{2} + H (J+2)^{3} (J+1)^3 \\
		& -\big[\gamma+\gamma_{D} (J+2) (J+1) + \gamma_{H} (J+2)^{2} \\
		& \times (J+1)^{2}\big] (J+2) + \big[\lambda + \lambda_{D} (J+2) (J+1) \\
		& + \lambda_{H} (J+2)^{2} (J+1)^{2}\big] \Bigg(\frac{2}{3}-\frac{2 (J+1)}{2 J+1}\Bigg)
	\end{split}
	\\
	\begin{split}
		H^{J-1, J+1} {=} & \left\langle N{=}J-1 | H | N{=}J+1 \right\rangle {=}
		H^{J+1, J-1} {=} \\
		& \big[\lambda + \lambda_{D} (J^{2} +J +1) + \lambda_{H} (J^{2} +J +1)^{2}\big] \\
		& \times \frac{2\sqrt{J(J+1)}}{2 J+1} ,
	\end{split}
\end{align}
where $\lambda$ and $\gamma$ are the spin-spin and spin-rotation constants, respectively. Subscripts $D$ and $H$ denote the first and second order centrifugal distortion constants. The constants $B$, $D$, and $H$ correspond to the rotational constant and its first and second order centrifugal distortion  terms, respectively. The measured values of these spectroscopic parameters for the electronic ground state $^{3}\Sigma^{-}_{g}$ of O\textsubscript{2} are listed in Tab.~\ref{tab2}, taken from Ref.~\cite{drouin2010}.

The evaluation of the above results involves the Clebsch-Gordan decomposition of the total angular momentum $\mathbf{J}$ into the rotation quantum number $\mathbf{N}$ and the total electron spin $\mathbf{S}$:
\begin{equation}\label{eq:cg}
	\left\vert N m_{N} , \, S m_{S}\right\rangle {=}  {\sum_{J{=}|N-S|, m_{J}{=}-J} ^{J{=}N+S, m_{J}{=}J}} \left\langle J m_{J} |N m_{N} , \, S m_{S} \right\rangle
	\left\vert J m_{J} \right\rangle ,
\end{equation}
where $\left\langle J m_{J} | N m_{N} , \, S m_{S} \right\rangle=\left\langle N m_{N} , \, S m_{S} |J m_{J} \right\rangle$ are the Clebsch-Gordan coefficients. This basis transformation allows for expressing the Hamiltonian matrix in the uncoupled representation, where the electron spin polarization is directly accessible.

\begin{table}
	\centering
		\begin{tabular}{c c}
			\hline\hline
			Parameter & Frequency in MHz  \\
			\hline 
			$B$ & $43100.44276$  \\  
			$D$ & $145.1271\times 10^{-3}$  \\ 
			$H$ & $49\times 10^{-9}$   \\
			$\lambda$ & $-59501.3438$    \\
			$\lambda_{D}$ & $-58.3680\times 10^{-3}$    \\
			$\lambda_{H}$ & $-290.8\times 10^{-9}$    \\
			$\gamma$ & $-252.58634$    \\
			$\gamma_{D}$ & $-243.42\times 10^{-6}$    \\
			$\gamma_{H}$ & $-1.46\times 10^{-9}$    \\
			\hline\hline			
		\end{tabular}
	\caption{Measured spectroscopic parameters for the electronic ground state $^{3}\Sigma^{-}_{g}$ of O\textsubscript{2}, from Ref.~\cite{drouin2010}.}\label{tab2}
\end{table}

\subsection{NO}

\begin{figure*}
	\includegraphics[width=1.0\textwidth]{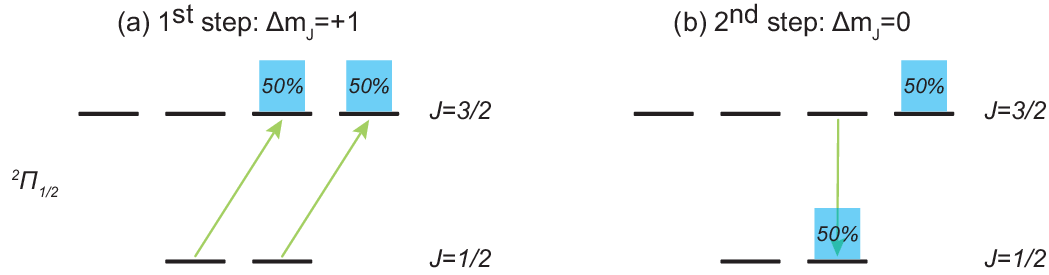}
	\caption{Excitation-polarization scheme for the $^{2}\Pi_{1/2}$ state of NO: (a) Transitions with $\Delta m_{J} = 1$: $\left\vert J = 1/2 \, m_{J} = - 1/2 \right\rangle \rightarrow\left\vert 3/2 \, 1/2 \right\rangle$ and $\left\vert 1/2 \, 1/2 \right\rangle\rightarrow\left\vert 3/2 \, 3/2 \right\rangle$. (b) Transition with $\Delta m_{J} = 0$: $\left\vert 3/2 \, 1/2 \right\rangle\rightarrow\left\vert 1/2 \, 1/2 \right\rangle$.}\label{figno}
\end{figure*}

Nitric oxide (NO) is a stable free radical with one unpaired electron and a doublet electronic ground state, $^{2}\Pi$. The energy separation between the lower $^{2}\Pi_{1/2}$ state and the upper $^{2}\Pi_{3/2}$ is about $123$~$\text{cm}^{-1}$~\cite{brown1972,saleck1992}. In this study, both nitrogen isotopes, \textsuperscript{14}N ($I=1$) and \textsuperscript{15}N ($I=1/2$), are considered following the excitation scheme illustrated in Fig.~\ref{figno}. As a result, the system is initially prepared in the states $\left\vert 1/2 \, 1/2 \right\rangle$ and $\left\vert 3/2 \, 3/2 \right\rangle$ (see Fig.~\ref{figno}(b)), with equal population and zero nuclear spin polarization. \rev{The corresponding electric dipole (E1) transitions occur at wavelengths of $\sim 2$~mm for \textsuperscript{14}N and $\sim 2.1$~mm for \textsuperscript{15}N in the ground vibrational state~\cite{wong2017}.} The corresponding Hamiltonian matrix elements (in frequency units), as given in Ref.~\cite{saleck1991}, are listed below:
\begin{align}
	\begin{split}
		& \left\langle I , ^{2}\Pi_{1/2} \, J^{\prime} ; F \pm | H |I, ^{2}\Pi_{1/2} \, J ; F \pm \right\rangle 
		= \\
		& \Bigg[ -\frac{1}{2} A_{eff} -\frac{1}{2} A_{D,eff} (X + 1) +B_{eff} (X + 1) \\
		&- D [(X + 1)^{2} + X] \pm \frac{1}{2} (-1)^{J-1/2 } \Big[[p+p_{D} (X+1)] \\
		& \times \bigg(J+\frac{1}{2}\bigg) +2[q+q_{D}(X+1)]\bigg(J+\frac{1}{2}\bigg)\Big]\Bigg] \delta_{J^{\prime}J}\\
		& + G(I, J^{\prime} , J, F) \Bigg[ (-1)^{J^{\prime}-1/2}
		\begin{pmatrix}
			J^{\prime} & 1 & J \\
			-\frac{1}{2} & 0 & \frac{1}{2}
		\end{pmatrix}
		\bigg(a-\frac{b+c}{2} \\
		& + \delta_{J^{\prime}J} 2 X C_{I}\bigg)
		\mp 
		\begin{pmatrix}
			J^{\prime} & 1 & J \\
			\frac{1}{2} & -1 & \frac{1}{2}
		\end{pmatrix}
		\frac{d + \delta_{J^{\prime}J} Xd_{D}}{\sqrt{2}}\Bigg] \\
		& + Q(I, J^{\prime} , J, F) (-1)^{J^{\prime}-1/2}
		\begin{pmatrix}
			J^{\prime} & 2 & J \\
			-\frac{1}{2} & 0 & \frac{1}{2}
		\end{pmatrix}
		\frac{eQq_{1}}{4}
	\end{split}
	\\
	\begin{split}
		& \left\langle I , ^{2}\Pi_{3/2} \, J^{\prime} ; F \pm | H |I, ^{2}\Pi_{3/2} \, J ; F \pm \right\rangle
		= \\
		& \Bigg[ \frac{1}{2} A_{eff} +\frac{1}{2} A_{D,eff} (X - 1) + B_{eff} (X - 1) \\
		& - D [(X - 1)^{2} + X] \pm \frac{1}{2} (-1)^{J-1/2 } q_{D}X\bigg(J+\frac{1}{2}\bigg)\Bigg] \delta_{J^{\prime}J}\\
		& + G(I, J^{\prime} , J, F) (-1)^{J^{\prime}-3/2}
		\begin{pmatrix}
			J^{\prime} & 1 & J \\
			-\frac{3}{2} & 0 & \frac{3}{2}
		\end{pmatrix}
		\bigg(a+\frac{b+c}{2} \\
		& + \delta_{J^{\prime}J} \frac{2}{3} X C_{I}\bigg) + Q(I, J^{\prime} , J, F) (-1)^{J^{\prime}-3/2}
		\begin{pmatrix}
			J^{\prime} & 2 & J \\
			-\frac{3}{2} & 0 & \frac{3}{2}
		\end{pmatrix} \\
		& \times \frac{eQq_{1}}{4}
	\end{split}
	\\
	\begin{split}
		& \left\langle I , ^{2}\Pi_{3/2} \, J^{\prime} ; F \pm | H |I, ^{2}\Pi_{1/2} \, J ; F \pm \right\rangle
		= \\
		& \Big[ - B_{eff} X^{1/2} + 2 D X^{3/2} \mp \frac{1}{2} \big[q + q_{D}(X+1)\big] \\
		& \times \bigg(J+\frac{1}{2}\bigg) X^{1/2}\Big] \delta_{J^{\prime}J} - G(I, J^{\prime} , J, F) (-1)^{J^{\prime}-3/2} \\
		&\times \begin{pmatrix}
			J^{\prime} & 1 & J \\
			-\frac{3}{2} & 1 & \frac{1}{2}
		\end{pmatrix} 
		\frac{b \pm \delta_{J^{\prime}J} (-1)^{J-1/2}\Big(J+\frac{1}{2}\Big)C_{I} ^{\prime}}{\sqrt{2}} \\
		& \mp Q(I, J^{\prime} , J, F)
		\begin{pmatrix}
			J^{\prime} & 2 & J \\
			\frac{3}{2} & -2 & \frac{1}{2}
		\end{pmatrix}
		\frac{eQq_{2}}{4\sqrt{6}} ,
	\end{split}
\end{align}
where $X = J (J + 1) - \frac{3}{4}$ and
\begin{align}
	\begin{split}
		G(I, J^{\prime} , J, F)  = & (-1)^{I + J^{\prime} + F} 
		\begin{Bmatrix}
			F & J^{\prime} & I \\
			1 & I & J
		\end{Bmatrix}\\
		& \times \sqrt{I(I+1)(2I+1)(2 J^{\prime} + 1)(2 J+1)} 
	\end{split}\\
	\begin{split}
		Q(I, J^{\prime} , J, F)  = & (-1)^{I + J^{\prime} + F}
		\begin{Bmatrix}
			F & J^{\prime} & I \\
			1 & I & J
		\end{Bmatrix} 
		\sqrt{(2J^{\prime} +1)}\\
		& \times \sqrt{(2J+1)\frac{(I+1)(2I+1)(2I+3)}{I (2 I - 1)}} .
	\end{split}
\end{align}

The total angular momentum $\mathbf{F}$ arises from the coupling of $\mathbf{J}$ and $\mathbf{I}$. \rev{The total angular momentum excluding nuclear spin, $\mathbf{J}$, is obtained by coupling $\mathbf{N}$ and $\mathbf{S}$ (see Eq.~(\ref{eq:cg})).} The signs $\pm$ in the bra-ket notation indicate the parity of the states. The experimental values of all spectroscopic constants are summarized in Tab.~\ref{tab3}. The above matrix elements yield a Hamiltonian in the so-called coupled representation. To compute observables such as the average rotational angular momentum projection and nuclear spin projection, a basis transformation is performed to express the Hamiltonian in the uncoupled representation. This allows direct evaluation of these quantities using a code similar to Sec.~\ref{sechd}. In practice, the Hamiltonian matrix size in the uncoupled basis is $72{\times} 72$ for \textsuperscript{14}NO and $48{\times} 48$ for \textsuperscript{15}NO.

\begin{table}
	\centering
		\begin{tabular}{c c c}
			\hline\hline
			\multirow{2}{*}{Parameter} & \multicolumn{2}{c}{Frequency in MHz} \\
			& \textsuperscript{14}NO & \textsuperscript{15}NO \\
			\hline
			$A_{eff}$ & $3691619$ & $3691683$ \\
			$A_{D,eff}$ & $5.497$ & $5.120$ \\
			$B_{eff}$ & $50847.7988$ & $49050.532$ \\
			$D$ & $0.164073$ & $0.1525$ \\
			$p$ & $350.37517$ & $337.9627$ \\
			$p_{D}$ & $0.000086$ & $0.00011$ \\
			$q$ & $2.83713$ & $2.64012$ \\
			$q_{D}$ & $0.000044$ & $0.0000379$ \\
			$a$ & $84.2155$ & $-118.143$ \\
			$b$ & $42.099$ & $-59.024$ \\
			$c$ & $-58.989$ & $82.725$ \\
			$d$ & $112.5972$ & $-157.9474$ \\
			$d_{D}$ & $0.00016$ & $-$ \\
			$C_{I}$ & $0.01242$ & $-0.01622$ \\
			$C_{I} ^{\prime}$ & $0.0039$ & $-0.0055$ \\
			$eQq_{1}$ & $-1.8581$ & $-$ \\
			$eQq_{2}$ & $23.153$ & $-$ \\
			\hline\hline
		\end{tabular}
	\caption{Measured spectroscopic constants for the electronic ground states of \textsuperscript{14}NO and \textsuperscript{15}NO, as reported in Ref.~\cite{saleck1991}.}\label{tab3}
\end{table}

\begin{figure*}
	\centering
	\includegraphics[width=1.0\textwidth]{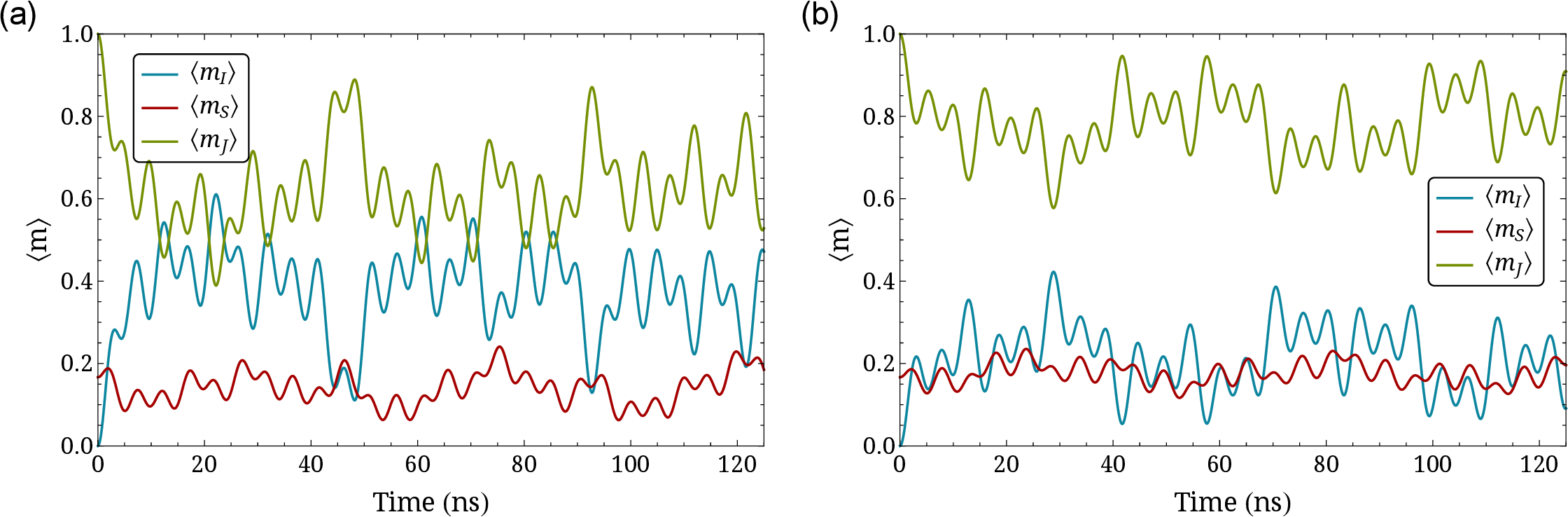}
	\caption{Time evolution of the average projections $\langle m_{J}\rangle$ (green), \rev{$\langle m_{S}\rangle$ (red),} and $\langle m_{I}\rangle$ (blue) for the molecules: (a) \textsuperscript{14}NO and (b) \textsuperscript{15}NO.}\label{fig5}
\end{figure*}

Figure~\ref{fig5} shows the time evolution of \rev{$\langle m_{J}\rangle$, $\langle m_{S}\rangle$, and $\langle m_{I}\rangle$} for the \textsuperscript{14}N (a) and \textsuperscript{15}N (b) isotopes. In the case of \textsuperscript{14}N, the nuclear polarization reaches $61.1\%$ at $22.2$~ns after excitation, whereas for \textsuperscript{15}N, it reaches $84.5\%$ at $28.8$~ns. At these specific times, the hyperfine beatings can be effectively frozen, thereby preserving the developed nuclear spin polarization.

\subsection{N\textsubscript{2}O}

Nitrous oxide (N\textsubscript{2}O) is a linear molecule with nitrogen as the central atom, making the two nitrogen nuclei distinguishable due to their nonidentical interactions. We label the outer nitrogen and central nitrogen nuclei with indices $1$ and $2$, respectively. For \textsuperscript{14}N\textsubscript{2}O, we consider the excitation sequence $\left\vert J = 0 \, m_{J} = 0 \right\rangle \rightarrow\left\vert 1 \, 1 \right\rangle \rightarrow\left\vert 2 \, 2 \right\rangle$, selected to achieve high polarization, as it involves two spin-$1$ nuclei. \rev{These electric dipole (E1) transitions correspond to wavelengths of $\sim 11.9$~mm and $\sim 6$~mm in the ground vibrational state~\cite{huang2025}.} 

The effective hyperfine Hamiltonian of \textsuperscript{14}N\textsubscript{2}O, expressed in frequency units and following the sign convention of Ref.~\cite{casleton1975} for the spin-rotation terms, is given by:
\begin{equation}\label{eq:n2o}
	\begin{split}
		H/h = & -M_{N_{1}} \mathbf{I_{1}} \cdot \mathbf{J} - M_{N_{2}} \mathbf{I_{2}} \cdot \mathbf{J} \\
		& +\frac{(eqQ)_{N_{1}}}{2 I_{1} (2 I_{1}-1) J (2 J -1)}\Big[3 (\mathbf{I_{1}} \cdot \mathbf{J})^2 + \frac{3}{2} (\mathbf{I_{1}} \cdot \mathbf{J}) \\ 
		& - \mathbf{I_{1}}^2 \mathbf{J}^2 \Big] +\frac{(eqQ)_{N_{2}}}{2 I_{2} (2 I_{2}-1) J (2 J -1)}\Big[3 (\mathbf{I_{2}} \cdot \mathbf{J})^2 \\
		& + \frac{3}{2} (\mathbf{I_{2}} \cdot \mathbf{J}) - \mathbf{I_{2}}^2 \mathbf{J}^2 \Big] +\frac{2 D_{NN}}{(2 J -1)(2 J + 3)} \\
		& \times \Big[\frac{3}{2} (\mathbf{I_{1}} \cdot \mathbf{J})(\mathbf{I_{2}} \cdot \mathbf{J}) +\frac{3}{2} (\mathbf{I_{2}} \cdot \mathbf{J})(\mathbf{I_{1}} \cdot \mathbf{J}) - \mathbf{I_{1}} \cdot \mathbf{I_{2}} \mathbf{J}^2 \Big].
	\end{split}
\end{equation}
The first two terms describe the spin-rotation interaction. The next two terms account for the electric quadrupole interactions of the two nitrogen nuclei, and the final term represents the dipole spin-spin interaction between $\mathbf{I_{1}}$ and $\mathbf{I_{2}}$. The experimentally reported hyperfine constants for \textsuperscript{14}N\textsubscript{2}O~\cite{casleton1975}, along with estimated values for \textsuperscript{15}N\textsubscript{2}O are listed in Tab.~\ref{tab4}. The values for \textsuperscript{15}N\textsubscript{2}O are obtained by substituting the \textsuperscript{14}N g-factors with those of \textsuperscript{15}N and omitting the quadrupole interaction terms, as the nuclear spin of \textsuperscript{15}N is $1/2$.

\begin{table}
	\centering
		\begin{tabular}{c c c}
			\hline\hline
			\multirow{2}{*}{Parameter} & \multicolumn{2}{c}{Frequency in kHz} \\
			& \textsuperscript{14}N\textsubscript{2}O & \textsuperscript{15}N\textsubscript{2}O \\
			\hline
			$M_{N_{1}}$ & $-2.35$ & $3.30$ \\
			$M_{N_{2}}$ & $-2.90$ & $4.07$ \\
			$(eqQ)_{N_{1}}$ & $-776.7$ & $-$ \\
			$(eqQ)_{N_{2}}$ & $-269.4$ & $-$ \\
			$D_{NN}$ & $0.436$ & $0.858$ \\
			\hline\hline
		\end{tabular}
	\caption{Hyperfine constants for \textsuperscript{14}N\textsubscript{2}O and \textsuperscript{15}N\textsubscript{2}O. The values for \textsuperscript{14}N\textsubscript{2}O are adapted from Ref.~\cite{casleton1975}, while those for \textsuperscript{15}N\textsubscript{2}O are estimated based on molecular similarity and the \textsuperscript{15}N g-factor.}\label{tab4}
\end{table}

\begin{figure*}
	\centering
	\includegraphics[width=1.0\textwidth]{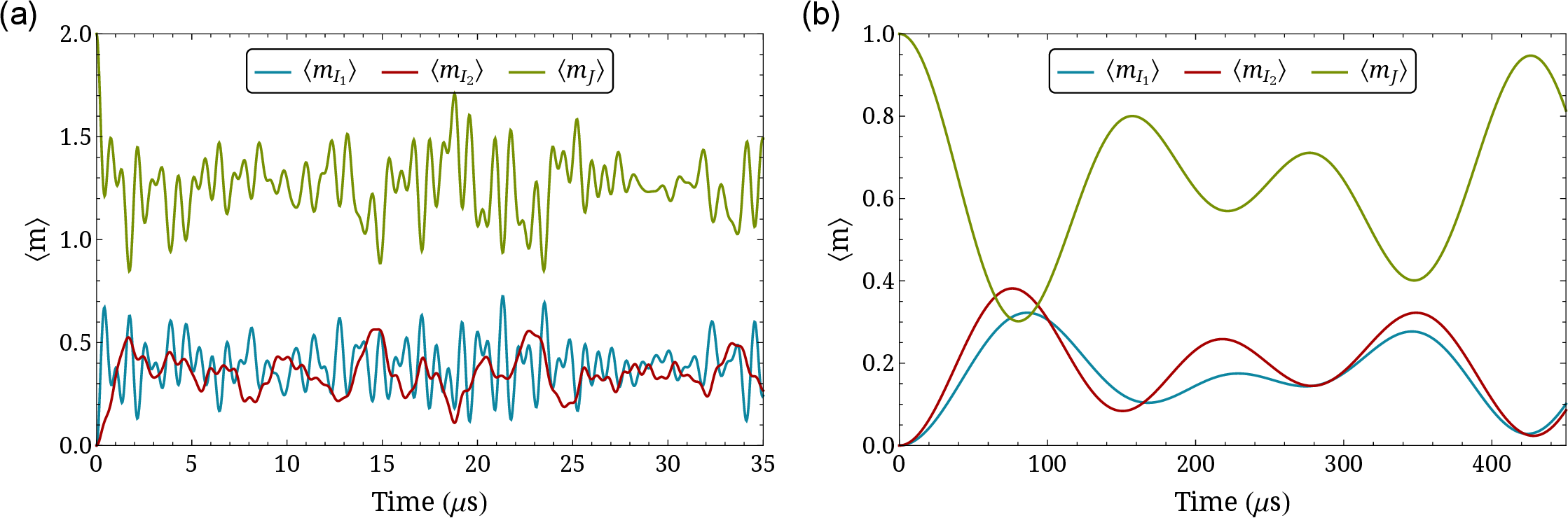}
	\caption{Time evolution of the average projections of rotational angular momentum $\langle m_{J}\rangle$ (green) and nitrogen nuclear spins $\langle m_{I_{1,2}}\rangle$ (blue and red) for the molecules: (a) \textsuperscript{14}N\textsubscript{2}O and (b) \textsuperscript{15}N\textsubscript{2}O.}\label{fig6}
\end{figure*}

As shown in Fig.~\ref{fig6}(a), the total nitrogen nuclear polarization  in \textsuperscript{14}N\textsubscript{2}O exceeds $57\%$ at times $1.7$~{\textmu}s and $23.5$~{\textmu}s. For \textsuperscript{15}N\textsubscript{2}O, a single excitation $\left\vert 0 \, 0 \right\rangle \rightarrow\left\vert 1 \, 1 \right\rangle$ \rev{(at $\sim 12.3$~mm~\cite{bauer1986})} is considered due to its different total nuclear spin. The resulting hyperfine beatings, depicted in Fig.~\ref{fig6}(b), indicate that the total nuclear polarization reaches $69.8\%$ at $80.1$~{\textmu}s after excitation. The hyperfine dynamics can be halted at the peak of nuclear polarization, ensuring its preservation.

\subsection{\textsuperscript{13}CO}

\begin{figure*}
	\centering
	\includegraphics[width=1.0\textwidth]{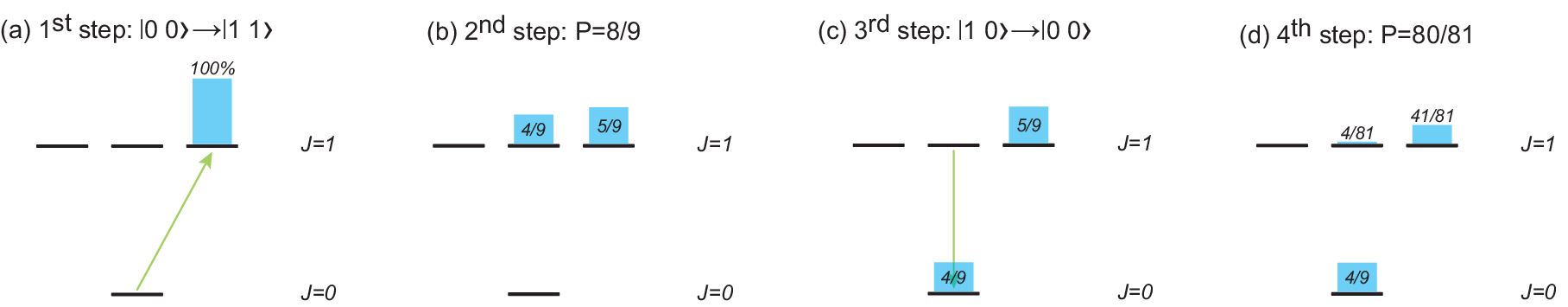}
	\caption{Excitation-polarization scheme for \textsuperscript{13}CO: (a) Excitation from the ground state $\left\vert 0 \, 0 \right\rangle$ to $\left\vert 1 \, 1 \right\rangle$. (b) Free evolution for $10.2$~{\textmu}s, during which the nuclear polarization reaches its maximum value of $8/9$. (c) De-excitation from $\left\vert 1 \, 0 \right\rangle$ to $\left\vert 0 \, 0 \right\rangle$. (d) Further free evolution for $10.2$~{\textmu}s, resulting in nuclear polarization of $80/81$.}\label{fig7}
\end{figure*}

Carbon-13 monoxide (\textsuperscript{13}CO) exhibits the simplest hyperfine structure among the molecules considered in this work. Its interaction is limited to the coupling between the nuclear spin $\mathbf{I}$ (with $I=1/2$) and the rotational angular momentum $\mathbf{J}$. The corresponding nuclear spin-rotation constant has been experimentally determined to be $32.63$~kHz~\cite{klapper2000}. Figure~\ref{fig7} illustrates a potential excitation scheme to achieve high nuclear polarization for the molecule. The system is initially excited from the ground state $\left\vert 0 \, 0 \right\rangle$ to $\left\vert 1 \, 1 \right\rangle$ \rev{with an electric dipole (E1) transition at $\sim 2.7$~mm~\cite{klapper2000}}, as shown in Fig.~\ref{fig7}(a). After a free evolution of $10.2$~{\textmu}s (see Fig.~\ref{fig8}(a)), the nuclear polarization reaches $8/9$, as shown in Fig.~\ref{fig7}(b). At this point, the populations of the $\left\vert 1 \, 0 \right\rangle$ and $\left\vert 1 \, 1 \right\rangle$ states are $4/9$ and $5/9$, respectively. Next, the $\left\vert 1 \, 0 \right\rangle$ state is de-excited, transferring $4/9$ of the total population to the ground state $\left\vert 0 \, 0 \right\rangle$ (see Fig.~\ref{fig7}(c)). Since the ground state exhibits no hyperfine beating, it retains its $100\%$ nuclear polarization. Allowing the system to evolve for an additional $10.2$~{\textmu}s results in population redistribution among the $J=1$ states, as illustrated in Fig.~\ref{fig7}(d). These states exhibit a nuclear polarization of $44/45$. Considering the fully nuclear-spin-polarized population of the ground state, the average nuclear polarization across the molecule reaches $80/81$.

\subsection{H\textsuperscript{32}S\textsuperscript{34}SH}

Hydrogen disulfide (H\textsubscript{2}S\textsubscript{2}) is a symmetric molecule that is an almost accidentally prolate  symmetric top based on its principal moments of inertia. When the sulfur nuclei are identical, the molecular symmetry restricts the total nuclear spin of the two hydrogen nuclei for different rotational levels, leading to ortho- and para-species associated with rotational levels of different symmetries~\cite{winnewisser1968}. Substituting one \textsuperscript{32}S with \textsuperscript{34}S (both isotopes have $I=0$) breaks this symmetry, allowing all possible proton spin combinations for every rotational level $J$ of H\textsuperscript{32}S\textsuperscript{34}SH (which has a natural abundance of $8\%$~\cite{behrend1990}). Since the molecule's total nuclear spin arises solely from the two protons ($I_{1,2} = 1/2$), a single excitation $\left\vert 0 \, 0 \right\rangle \rightarrow \left\vert 1 \, 1 \right\rangle$ can efficiently transfer rotational angular momentum polarization to the nuclear spins. \rev{The electric dipole (E1) transition for this excitation in the state discussed below ($1_{0 1}$) occurs at $\sim 22.1$~mm~\cite{winnewisser1991}.}

\begin{figure*}
	\centering
	\includegraphics[width=1.0\textwidth]{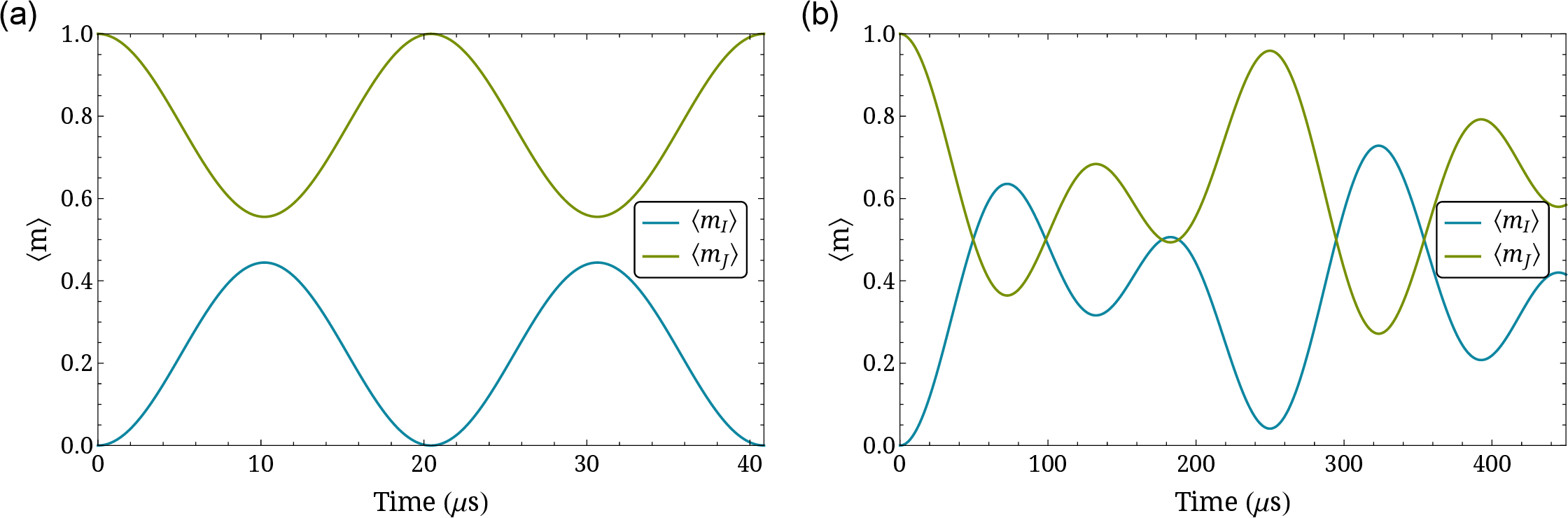}
	\caption{Time evolution of the average projections of rotational angular momentum $\langle m_{J}\rangle$ (green) and nuclear spin $\langle m_{I}\rangle$ (blue) for the molecules: (a) \textsuperscript{13}CO and (b) H\textsuperscript{32}S\textsuperscript{34}SH.}\label{fig8}
\end{figure*}

The effective hyperfine Hamiltonian includes nuclear spin-rotation interactions and the dipole spin-spin interaction between the two protons, and can be written in frequency units as:
\begin{equation}\label{eq:h2s2}
	\begin{split}
		H/h = &  C_{J} (\mathbf{I_{1}} \cdot \mathbf{J} + \mathbf{I_{2}} \cdot \mathbf{J}) +\frac{D_{J}}{J (2 J -1)}\Big[\frac{3}{2} (\mathbf{I_{1}} \cdot \mathbf{J})(\mathbf{I_{2}} \cdot \mathbf{J}) \\
		& +\frac{3}{2} (\mathbf{I_{2}} \cdot \mathbf{J})(\mathbf{I_{1}} \cdot \mathbf{J}) - \mathbf{I_{1}} \cdot \mathbf{I_{2}} \mathbf{J}^2 \Big],
	\end{split}
\end{equation}
where only the dipolar coupling constant, $D_{J}$ can be reliably calculated from molecular geometry~\cite{winnewisser1991}. The spin-rotation constant $C_{J}$ is estimated based on~\cite{saleck1995,saleck1995b,cazzoli2014}, providing a plausible order of magnitude in the absence of direct measurements. For the $1_{0 1}$ level, we estimate $C_{1_{0 1}} {\sim} -4$~kHz and calculate $D_{1_{0 1}} = 0.11$~kHz. Using the same methodology as for the other molecules, we simulate the resulting polarization beatings, as shown in Fig.~\ref{fig8}(b). The total nuclear polarization reaches ${\sim}64\%$ and ${\sim}73\%$ at ${\sim}72$~{\textmu}s and ${\sim}323$~{\textmu}s, respectively.

\section{Conclusion}
\label{conclusion}

\rev{We have presented a theoretical investigation of microwave and IR excitation schemes for the production of spin-polarized small molecules, including HD, DT, O\textsubscript{2}, NO, N\textsubscript{2}O, CO, and H\textsubscript{2}S\textsubscript{2}. We have analyzed the polarization dynamics induced by coherent rotational excitation and quantified the achievable nuclear spin polarization under experimentally relevant conditions. Our simulations indicate that, in principle, high degrees of polarization and substantial production rates could be achieved with intense radiation sources. The present work establishes the theoretical basis and identifies the parameter regimes required for a future experimental realization of the proposed schemes.}


\section*{Acknowledgments}
	This work was partially supported by the Hellenic Foundation for Research and Innovation (HFRI) and the General Secretariat for Research and Technology (GSRT), under grant agreement No.~HFRI-FM17-3709 (project NUPOL). CSK acknowledges funding from the Deutsche Forschungsgemeinschaft (DFG, German Research Foundation)-533904660.

%


\end{document}